# Finite-Size Effect Induced Spatial-Spectral Mode Splitting in Membrane Metasurfaces

Chih-Zong Deng,[1] Mu-Hsin Chen,[1] Chun-Hao Chiang,[1] Jui-Han Fu,[2] Vincent Tung,[2] Masanobu Iwanaga[1], and Ya-Lun Ho[1*]

[1]Research Center for Electronic and Optical Materials, National Institute for Materials Science (NIMS), Tsukuba, Ibaraki, Japan

[2]Department of Chemical System Engineering, School of Engineering, The University of Tokyo, Tokyo, Japan

Corresponding author: HO.Ya-Lun@nims.go.jp

## Abstract

This work reports the spatial-spectral engineering and finite-size quantization of optical modes within a triangular-lattice silicon nitride membrane metasurface. Truncating the lattice into a finite square cavity breaks translational symmetry and lifts modal degeneracy, splitting optical modes into discrete cavity-envelope sub-modes. High-resolution photoluminescence (PL) scanning reveals distinct spatial field distributions. The corner-localized sub-mode features the highest Q-factor due to multipolar far-field destructive interference, whereas the core-localized sub-mode exhibits strong radiative coupling. PL mapping reveals a symmetric, four-fold clover-like wavelength arrangement. These results demonstrate that boundary-induced deterministic symmetry can override underlying lattice characteristics, offering a robust strategy for precise spatial-spectral tailoring of light-matter interactions at the nanoscale.

## Introduction

Dielectric metasurfaces and photonic crystal slabs have emerged as a promising platform for manipulating light-matter interactions at the subwavelength scale. By structuring high-index dielectrics, researchers can precisely tailor the dispersion and localization of electromagnetic fields, enabling photonic applications such as low-threshold lasers[1-5], enhanced non-linear optical devices[6, 7], and highly sensitive sensors[8, 9]. Central to the performance of these applications is the engineering of resonant architectures capable of supporting high quality factor modes. Within this context, various resonant phenomena, including guided mode resonances[10] and bound states in the continuum (BIC)[11], have attracted intense interest due to their theoretical ability to achieve extreme field confinement and minimal radiative loss within a planar geometry. Conventionally, the optical characteristics of these high-quality resonances are analyzed under the ideal assumption of an infinite periodic lattice, where the modal behaviors are strictly dictated by the spatial point-group symmetry and the Bloch wavefunctions of the constituent unit cells.

To maximize the field confinement and quality factors of these optical resonances, freestanding membrane configurations have emerged as a promising architectural strategy to overcome the intrinsic limitations of conventional substrate-supported structures[3-6, 12-17]. In typical substrate-supported geometries, the presence of an underlying substrate inherently breaks out-of-plane vertical symmetry and introduces radiative leakage pathways, which severely weakens electromagnetic field localization. By entirely removing the substrate material, suspended nanomembrane structures restore a perfectly symmetric vertical refractive index profile, thereby eliminating substrate-induced radiation losses. This suspended configuration establishes an exceptional refractive index contrast between the high-index dielectric core and the surrounding air claddings, securely trapping electromagnetic energy within the thin-film plane and minimizing out-of-plane radiative attenuation. Furthermore, the absence of substrate-related perturbations supports robust in-plane and out-of-plane field localization, creating highly confined surface fields that significantly maximize light-matter interactions. This dielectric environment not only ensures strong integration compatibility but also enhances spatial mode overlap with active functional layers.

However, practical micro- and nanophotonic platforms inherently demand finite geometric dimensions to successfully integrate into operational chip architectures. Truncating an infinite periodic lattice into a finite spatial boundary inevitably disrupts the ideal translational symmetry of the crystal. This finite-size restriction imposes a severe constraint on the modal landscape, forcing the continuous flat bands of the lattice to discretize into localized, quantized states[3-5]. Specifically, the boundary-induced symmetry breaking lifts the modal degeneracy of high-order resonances at the Γ point, which manifests as complex multi-peak splitting and pronounced spatial-spectral modulation in experimental collection. Consequently, understanding how these finite boundaries discretize modes into distinct cavity-envelope patterns, and deciphering the exact relationship between finite cavity geometries and quality factor evolution, represents a crucial physical challenge in advancing metasurface engineering toward deterministic light manipulation. Although previous studies have explored the tuning of optical resonances using microcavity structures[3], experimental results relationship between the spatial far-field patterns of these split sub-modes and their quality factor evolution still limited. Elucidating the behaviors of sub-mode is essential for achieving deterministic control over light-matter interactions at the nanoscale.

In this systematic study, we investigated the splitting of optical modes in a triangular-lattice membrane metasurface via geometric thickness engineering and boundary confinement. We first analyze the near-field multipolar characteristics of various optical resonances through photonic band-structure calculations, distinguishing radiative leaky modes from symmetry-protected BICs. Using high-resolution PL scanning techniques, we systematically dissect the spatial field distributions and spectral evolution of the discrete sub-modes under finite square boundary constraints. This work not only elucidates how boundary-induced deterministic geometric symmetry governs and tailors the underlying lattice characteristics, but also opens up promising avenues for precise spatial-spectral engineering of photonic modes at the nanoscale.

## Results and Discussion

Fig.1 shows the mode analysis of the SiN membrane metasurface. The photonic band structure was calculated across the Brillouin zone of the triangular lattice under both transverse-electric (TE) and transverse-magnetic (TM) polarizations, as displayed in Fig. 1(a). For a membrane thickness of 160 nm, seven distinct, dispersive resonant bands, designated from A to G, are clearly resolved within the simulated spectral window at Γ point. To identity the multipolar characteristics of these optical resonances, the out-of-plane field distributions ($E_z$ for TM modes and $H_z$ for TE modes) were calculated at the Γ point, as shown in Fig. 1(b). The spatial field distributions reveal a diverse set of multipolar behaviors, including monopole, dipole, quadrupole, and hexapole distributions, which are tightly dictated by the inherent C6 symmetry of the hexagonal lattice. Specifically, modes A and C are identified as TM- and TE-polarized dipole modes, respectively. Due to their symmetry matching with the radiative continuum of free space, these two states act as conventional leaky modes characterized by strong radiation leakage. Consequently, they exhibit relatively low Q-factors ($10^2$ for mode A and $10^1$ for mode C at Γ point).

In contrast, the remaining five optical resonances including the TM quadrupole (mode B), TM hexapole (mode D), TE monopole (mode E), TE quadrupole (mode F), and TE hexapole (mode G) are strictly decoupled from the free-space continuum at the Γ point due to their specific spatial symmetries under the C6 point group. These states manifest as symmetry-protected BICs, where vertical radiative coupling is theoretically prohibited by the symmetry mismatch between the metasurface eigenmodes and external plane waves. Consequently, at the Γ point these BIC modes feature vanishing radiative decay channels and infinite Q-factors, providing an ideal platform for engineering strong light–matter interactions.

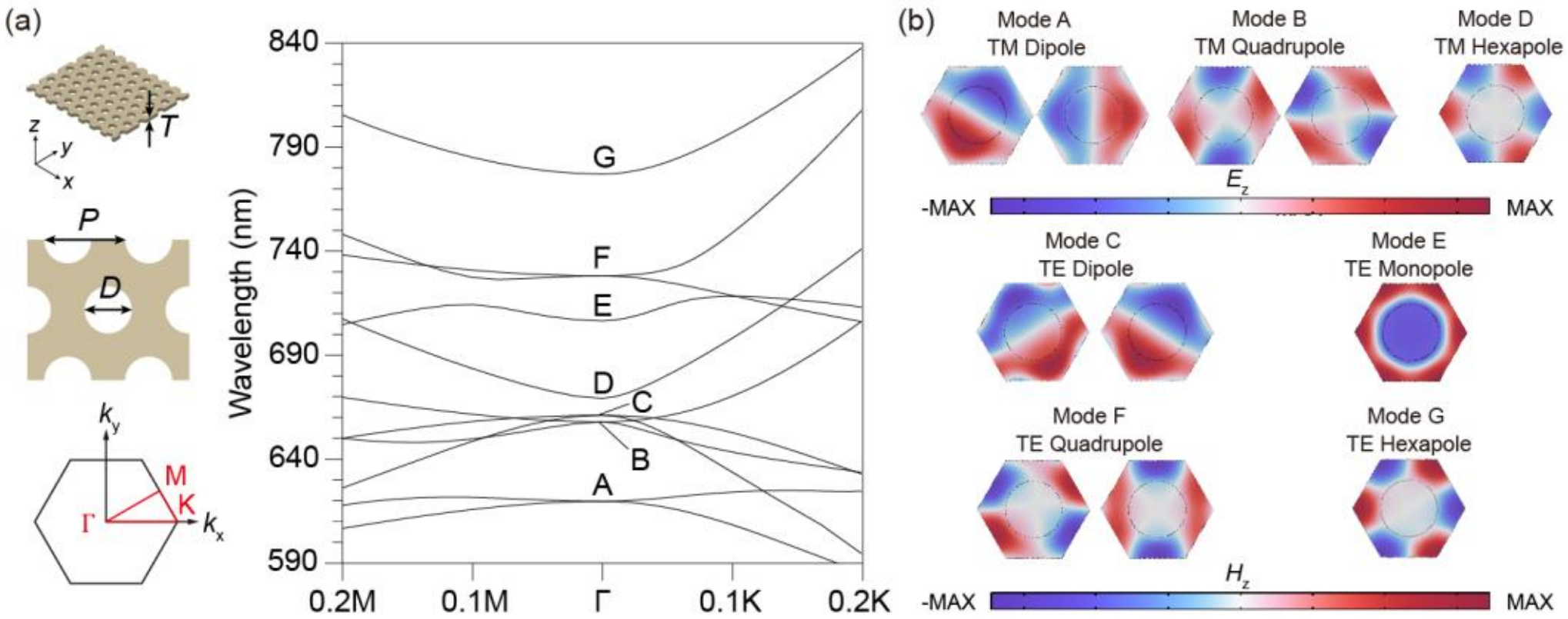

FIG. 1. (a) Schematic diagram of the hole-array SiN membrane metasurface and its corresponding photonic band structure, featuring a lattice period $P$ = 600 nm, hole diameter $D$ = 350 nm, and membrane thickness $T$ = 160 nm. (b) Simulated field distributions of the out-of-plane electric ($E_z$) and magnetic ($H_z$) field components for the TM and TE modes, respectively, calculated at the Γ point.

To experimentally validate the numerically predicted optical resonances, PL spectroscopy was performed on the membrane metasurfaces (Fig. 2). Figs. 2a and 2c display the localized PL spectra acquired from the metasurfaces with membrane thicknesses of 200 nm and 160 nm, respectively. The black and red curves represent the localized spectra measured at different spatial coordinates on the metasurface, corresponding to the locations indicated by the black and red dots in the inset OM images. Several emission peaks are clearly resolved, matching the characteristic resonances from mode A to mode G predicted by the band structure calculation in Fig. 1. Notably, because the intrinsic background emission of the SiNx membrane drastically diminishes at wavelengths longer than 800 nm (as detailed in Fig. SX), the resonance of mode G is initially suppressed and remains barely observable in the 200-nm-thick sample shown in Fig. 2a. To overcome this spectral limitation and thoroughly investigate the emission characteristics of mode G, the sample was thinned down to a thickness of 160 nm by carefully controlled RIE process. This geometric thinning induces a pronounced blueshift across all optical modes, successfully shifting the resonance of mode G into the higher-efficiency emission window of the SiNx membrane. Consequently, a prominent multi-peak profile for mode G clearly emerges in the 160-nm-thick sample spectrum in Fig. 2c.

Due to the finite-size effect inherent to the bounded cavity of the metasurface (with a lateral footprint of 25 x 20 um), a clear spatial-spectral modulation is manifested. The localized spectra collected from different positions vary substantially, suggesting that each split resonance possesses a distinct spatial emission pattern. Specifically, the truncation of the lattice by finite boundaries breaks the ideal symmetry, causing mode E to split into discrete sub-modes E1 and E2, mode F into F1 and F2, and mode G into an even finer cluster of discrete states. To visualize these spatial characteristics, spatial scanning was performed to map the PL emission intensity profiles across the metasurface, as shown in Figs. 2b and 2d for the 200-nm and 160-nm metasurfaces, respectively. Remarkably, despite the spectral blueshift induced by the change in membrane thickness, the spatial emission fingerprints for each individual mode remain exceptionally stable and look nearly identical between the two different thicknesses. This robust preservation of spatial symmetry demonstrates that the spatial profiles of these cavity-envelope states are rigidly dictated by the deterministic geometric boundary, serving as a powerful visual metric for unambiguous mode identification.

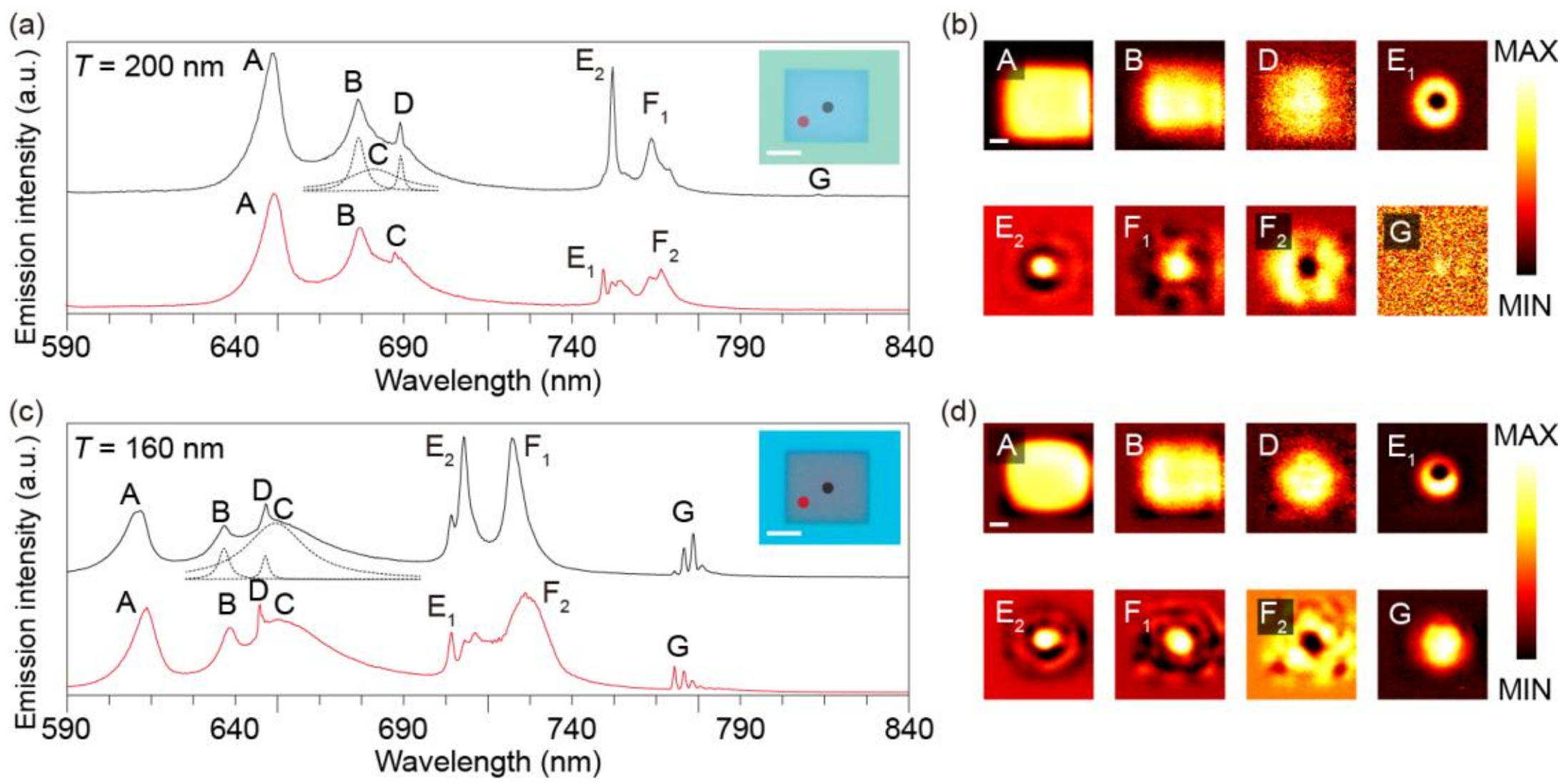

FIG. 2. PL emission spectra measured at different locations for metasurface thicknesses of (a) 200 nm and (c) 160 nm. The spectra are deconvoluted into individual Lorentzian peaks (dashed curves). Inset: Optical microscope (OM) images of the corresponding metasurfaces, where black and red spots indicate the respective measurement locations. The scale bar represents 100 nm. (b, d) Spatial emission intensity distributions of the optical resonances labeled A–G in (a) and (c). The scale bar represents 5 μm.

To systematically analyze the behavior of mode G, the metasurface thickness was thinned down to 92 nm via a carefully controlled RIE process. This geometric tailoring shifts the resonance of mode G into spectral overlap with the intrinsic PL emission peak of the SiNx membrane. As shown in Fig. 3(a), the PL spectrum of the 92-nm-thick membrane metasurface reveals a prominent resonance feature designated as mode G at approximately 700 nm, where multiple split sub-modes begin to emerge clearly. To elucidate the exact fine spectral features of this multi-peak splitting behavior, we performed high-resolution PL scanning across localized regions of the metasurface, as illustrated in Fig. 3(b). By positioning the probe spot at distinct spatial coordinates, we successfully resolved four sharp, discrete resonance peaks labeled G1, G2, G3, and G4, spanning from 696.45 nm to 701.85 nm.

Crucially, the relative emission intensities of these four co-existing spectra vary depending on the exact measurement location. For instance, when probing near the top-right corner along the diagonal direction of the square boundary, the shortest-wavelength resonance G1 dominates the spectrum, exhibiting an exceptionally sharp profile with the highest quality factor. In contrast, when shifting the probe spot along the orthogonal y- or x-directions, the intermediate sub-modes G2 and G3 dominate. These two modes are spectrally closely spaced and nearly merged; notably, their resonance wavelengths are shorter than that of the central mode (G4) but longer than the diagonal-associated resonance (G1). This trend reflects the spatial quantization within the square boundary, where the effective cavity confinement along the principal axes dictates intermediate resonant states between the diagonal corners and the geometric center. Meanwhile, probing the exact geometric center of the metasurface isolates the longest-wavelength sub-mode G4, which manifests a much broader profile with the lowest Q-factor. This strong position-dependent spectral evolution provides definitive experimental evidence of spatial-spectral splitting within the cavity.

To visually correlate these discrete spectral peaks with their PL emission distributions, we obtained the spatial PL emission intensity profiles (mode patterns) for each sub-mode via high-resolution scanning across the metasurface. As illustrated in Fig. 3(c), the four split resonances exhibit remarkably distinct spatial field configurations. Mode G1 displays a well-defined four-lobed pattern with intense field localization near the outer corners of the square metasurface. In contrast, modes G2 and G3 manifest as dimeric, two-lobed configurations aligned along orthogonal axes, while mode G4 presents a highly centralized hotspot pattern concentrated at the geometric core. Even though the underlying metasurface possesses a triangular lattice symmetry, its truncation by the finite-size square boundary profile discretizes the modes into these distinct cavity-envelope patterns.

To obtain a comprehensive overview of this sub-mode distribution, we performed a mapping of the dominant PL emission peak wavelengths across the entire sample area. As demonstrated in Fig. 3(d), the resulting color-coded wavelength map unveils a highly symmetric, four-fold clover-like spatial arrangement that encapsulates the boundaries of the split modes (G1 to G4). The central domain is characterized by longer wavelengths corresponding to the centralized G4 mode, while the surrounding quadrants transition into distinct color zones that mirror the orthogonal and corner symmetries observed in the individual intensity patterns of G1 to G3. This clear spatial-spectral mapping unambiguously confirms that the multi-peak splitting of mode G is governed by the deterministic geometric symmetry of the membrane metasurface, opening up promising avenues for precise spatial-spectral tailoring of light-matter interactions at the nanoscale.

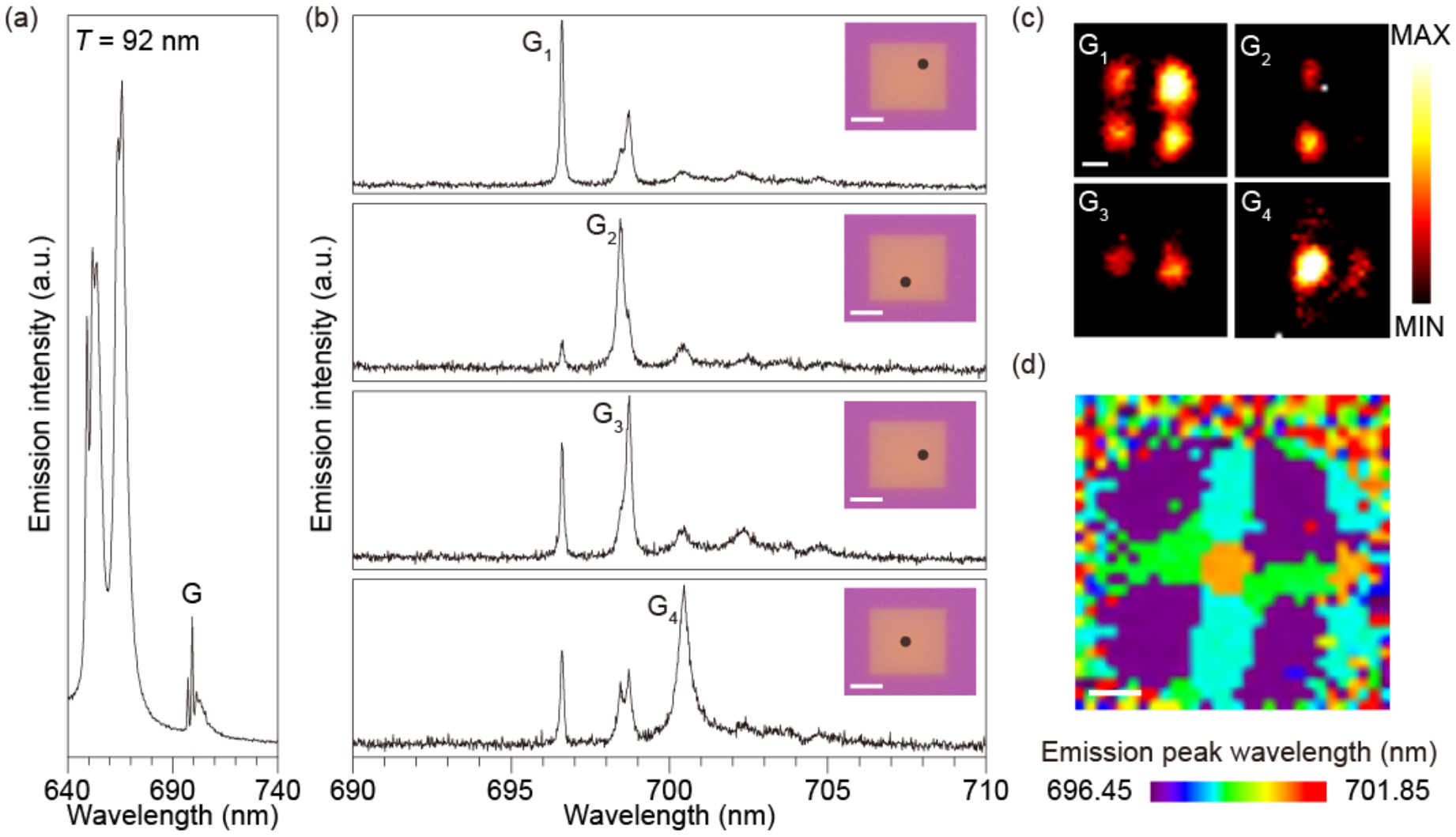

FIG. 3. (a) PL spectrum of the membrane metasurface with a thickness of 92 nm. (b) High-resolution PL spectrum focusing on the mode G region. Inset: OM image of the metasurface, where black spots indicate the measurement locations. (c) Spatial emission intensity

distributions of the optical resonances labeled $G_1$–$G_4$ in (b). (d) Spatially resolved mapping of the emission peak wavelength for mode G.

In summary, the spatial-spectral engineering and finite-size quantization of symmetry-protected BICs within a triangular-lattice SiNx membrane metasurface have been demonstrated. While infinite-periodic band calculations identify mode G at the Γ point as a hexapolar BIC with a infinite Q-factor, its experimental result is initially hindered by the weak background emission of SiNx at longer wavelengths. To overcome this, controlled RIE was utilized to thin the membrane to 92 nm, blueshifting the mode G resonance into the high-efficiency emission window of SiNx at approximately 700 nm. Truncating the lattice into a finite 25 x 20 um square cavity breaks translational symmetry, splitting mode G into four discrete cavity-envelope states (G1 to G4). High-resolution PL scanning reveals distinct spatial field distributions: the corner-localized G1 mode features the highest Q-factor due to multipolar far-field destructive interference, whereas the core-localized G4 mode exhibits strong radiative coupling. Additionally, spectral mapping reveals a symmetric, four-fold clover-like wavelength arrangement. These results demonstrate that boundary-induced deterministic symmetry can override underlying lattice characteristics, offering a robust strategy for precise spatial-spectral tailoring of light-matter interactions.

## Acknowledgments

This work was supported by JSPS KAKENHI under Grant Numbers JP23K26155, JP25KF0083, and JP25H01614. Part of this work was also supported by the Ministry of Education, Culture, Sports, Science and Technology (MEXT) under the MEXT–Quantum Leap Flagship Program (Grant Number JPMXS0118067246) and the Advanced Research Infrastructure for Materials and Nanotechnology in Japan (ARIM) (Proposal Number JPMXP1225NM5090).

## Data availability statement

The data that support the findings of this study are available in the supplementary material of this article.

## References

(1) Hwang, M. S.; Lee, H. C.; Kim, K. H.; Jeong, K. Y.; Kwon, S. H.; Koshelev, K.; Kivshar, Y.; Park, H. G., Ultralow-threshold laser using super-bound states in the continuum. *Nat. Commun.* **2021,** *12* (1), 4135.
(2) Xing, D.; Chen, M. H.; Wang, Z.; Deng, C. Z.; Ho, Y. L.; Lin, B. W.; Lin, C. C.; Chen, C. W.; Delaunay, J. J., Solution-Processed Perovskite Quantum Dot Quasi-BIC Laser from Miniaturized Low-Lateral-Loss Cavity. *Advanced Functional Materials* **2024,** *34* (26), 2314953.
(3) Contractor, R.; Noh, W.; Redjem, W.; Qarony, W.; Martin, E.; Dhuey, S.; Schwartzberg, A.; Kante, B., Scalable single-mode surface-emitting laser via open-Dirac singularities. *Nature* **2022,** *608* (7924), 692-698.
(4) Ren, Y.; Li, P.; Liu, Z.; Chen, Z.; Chen, Y.-L.; Peng, C.; Liu, J., Low-threshold nanolasers based on miniaturized bound states in the continuum. *Sci. Adv.* **2022,** *8*, eade8817.

(5) Zhong, H.; Yu, Y.; Zheng, Z.; Ding, Z.; Zhao, X.; Yang, J.; Wei, Y.; Chen, Y.; Yu, S., Ultra-low threshold continuous-wave quantum dot mini-BIC lasers. *Light Sci Appl* **2023,** *12* (1), 100.
(6) Konishi, K.; Akai, D.; Mita, Y.; Ishida, M.; Yumoto, J.; Kuwata-Gonokami, M., Circularly polarized vacuum ultraviolet coherent light generation using a square lattice photonic crystal nanomembrane. *Optica* **2020,** *7* (8).
(7) Bernhardt, N.; Koshelev, K.; White, S. J. U.; Meng, K. W. C.; Froch, J. E.; Kim, S.; Tran, T. T.; Choi, D. Y.; Kivshar, Y.; Solntsev, A. S., Quasi-BIC Resonant Enhancement of Second-Harmonic Generation in WS(2) Monolayers. *Nano Lett* **2020,** *20* (7), 5309-5314.
(8) Richter, F. U.; Sinev, I.; Zhou, S.; Leitis, A.; Oh, S. H.; Tseng, M. L.; Kivshar, Y.; Altug, H., Gradient High-Q Dielectric Metasurfaces for Broadband Sensing and Control of Vibrational Light-Matter Coupling. *Adv. Mater.* **2024,** *36* (25), e2314279.
(9) Wang, J.; Kühne, J.; Karamanos, T.; Rockstuhl, C.; Maier, S. A.; Tittl, A., All-Dielectric Crescent Metasurface Sensor Driven by Bound States in the Continuum. *Advanced Functional Materials* **2021,** *31* (46), 2104652.
(10) Huang, L.; Jin, R.; Zhou, C.; Li, G.; Xu, L.; Overvig, A.; Deng, F.; Chen, X.; Lu, W.; Alu, A.; Miroshnichenko, A. E., Ultrahigh-Q guided mode resonances in an All-dielectric metasurface. *Nat Commun* **2023,** *14* (1), 3433.
(11) Hsu, C. W.; Zhen, B.; Stone, A. D.; Joannopoulos, J. D.; Soljačić, M., Bound states in the continuum. *Nat. Rev. Mater.* **2016,** *1* (9), 1-13.
(12) Adi, W.; Rosas, S.; Beisenova, A.; Biswas, S. K.; Mei, H.; Czaplewski, D. A.; Yesilkoy, F., Trapping light in air with membrane metasurfaces for vibrational strong coupling. *Nat. Commun.* **2024,** *15* (1), 10049.
(13) Ho, Y. L.; Fong, C. F.; Wu, Y. J.; Konishi, K.; Deng, C. Z.; Fu, J. H.; Kato, Y. K.; Tsukagoshi, K.; Tung, V.; Chen, C. W., Finite-Area Membrane Metasurfaces for Enhancing Light-Matter Coupling in Monolayer Transition Metal Dichalcogenides. *ACS Nano* **2024,** *18* (35), 24173-24181.
(14) Deng, C. Z.; Shi, S.; Chiang, C. H.; Chen, M. H.; Liu, S.; Sakurai, H.; Fu, J. H.; Konishi, K.; Iwanaga, M.; Tung, V.; Ho, Y. L., Freestanding Polymer Metasurface Supporting Higher-Order Optical Resonances for Strong Field Enhancement in TMD Monolayers. *Small* **2026,** *22* (35), e13320.
(15) Deng, C. Z.; Chiang, C. H.; Shi, S.; Fu, J. H.; Wu, Y. J.; Konishi, K.; Tung, V.; Chen, C. W.; Ho, Y. L., Atomic-Scale Light Coupling Control in Ultrathin Photonic Membranes. *Advanced Functional Materials* **2026**, e24286.
(16) Hsieh, F. L.; Deng, C. Z.; Huang, S. K.; Liu, T. H.; Chen, M. H.; Chiang, C. H.; Lee, C. L.; Lai, M. H.; Fu, J. H.; Tung, V.; Chang, Y. M.; Chen, C. W.; Ho, Y. L., Large-Area Photonic Membranes Achieving Uniform and Strong Enhancement of Photoluminescence and Second-Harmonic Generation in Monolayer WSe(2). *Small Methods* **2026,** *10*, e01693.
(17) Ho, Y. L.; Chen, M. H.; Liu, T. H.; Hsieh, F. L.; Chiang, C. H.; Deng, C. Z.; Lai, M. H.; Shiue, J.; Liu, S.; Sakurai, H.; Fu, J. H.; Konishi, K.; Tung, V.; Chang, Y. M.; Chen, C. W.; Huang, S. K., Graphene-Scaffolded Ultrathin Perovskite Nanocrystal Films for Amplifying Energy Localization via Dual-Mode Nonhybridizing Quasi-BICs. *Nano Lett* **2026,** *26* (13), 4439-4448.